\documentstyle[aps,pra]{revtex}
\begin{document}
\draft

\newcommand{\uu}[1]{\underline{#1}}
\newcommand{\pp}[1]{\phantom{#1}}
\newcommand{\be}{\begin{eqnarray}}
\newcommand{\ee}{\end{eqnarray}}
\newcommand{\ve}{\varepsilon}
\newcommand{\vs}{\varsigma}
\newcommand{\Tr}{{\,\rm Tr\,}}
\newcommand{\pol}{\frac{1}{2}}

\title{
Reducible field quantization (II): Electrons
}
\author{Marek~Czachor}
\address{
Katedra Fizyki Teoretycznej i Metod Matematycznych\\
Politechnika Gda\'{n}ska,
ul. Narutowicza 11/12, 80-952 Gda\'{n}sk, Poland
}
\maketitle

\begin{abstract}
Quantization of free Dirac fields is formulated in terms of a reducible representation of CAR. Similarly to the bosonic case we arrive at field operators which are indeed operators and not operator valued distributions. Observables such as 4-momentum and charge can be defined without any need of normal ordering.
\end{abstract}


\section{Introduction: Jordan-Wigner-type construction of CAR}

The paper continues the program of reducible field quantization of electrodynamics \cite{1,2,3}. The experience with quantization of electromagnetic fields shows that many results are easiest to prove if one employs properties of a concrete representation. For this reason we begin with an explicit construction of the 
``$N$-oscillator" reducible representation of the algebra of canonical anti-commutation relations (CAR).  

The departure point is the Jordan-Wigner construction of CAR \cite{JW,PG}. We need four operators corresponding to negatons and positons \cite{electron}, with two spin degrees of freedom for each of them. We define 
\be
b_-
&=&
\sigma_-\otimes \bbox 1\otimes \bbox 1\otimes \bbox 1,\quad
b_+
=
-\sigma_3\otimes \sigma_-\otimes \bbox 1\otimes \bbox 1,\quad
d_-
=
\sigma_3\otimes \sigma_3\otimes \sigma_-\otimes\bbox 1,\quad
d_+
=
-\sigma_3\otimes \sigma_3\otimes \sigma_3\otimes \sigma_-,\\
{\bbox I}
&=&
\bbox 1\otimes \bbox 1\otimes \bbox 1\otimes \bbox 1.\label{CAR}
\ee
Here $\bbox 1$ is the $2\times 2$ identity matrix, $\sigma_\pm=(\sigma_1\pm i\sigma_2)/2$, and $\sigma_k$ are the Pauli matrices.
The operators satisfy CAR 
\be
\{b_s,b_s^{\dag}\}
&=&
\{d_s,d_s^{\dag}\}
=\bbox I
\ee
and all the remaining anti-commutators vanish. We shall also need the matrix 
\be
\bbox I_0
&=&
\sigma_3\otimes\sigma_3\otimes\sigma_3\otimes \sigma_3
\ee
which anti-commutes with $b_s$ and $d_s$. For any
$
A=\left(
\begin{array}{cc}
A_{--} & A_{-+}\\
A_{+-} & A_{++}
\end{array}
\right),
$
and $B=e^A$, denoting
\be
b^{\dag}Ab
=
\sum_{s,s'=\pm}b_{s}^{\dag}A_{ss'}b_{s'},\quad
d^{\dag}Ad
=
\sum_{s,s'=\pm}d_{s}^{\dag}A_{ss'}d_{s'},
\ee
we find \cite{tensor}
\be
e^{b^{\dag}Ab}
=
\left(
\begin{array}{cccc}\det B & 0 & 0 & 0\\
0 & B_{--} & B_{-+} & 0\\
0 & B_{+-} & B_{++} & 0\\
0 & 0 & 0 & 1
\end{array}
\right)
\otimes \bbox 1 \otimes \bbox 1,\quad
e^{d^{\dag}Ad}
=
\bbox 1 \otimes \bbox 1 \otimes 
\left(
\begin{array}{cccc}
\det B & 0 & 0 & 0\\
0 & B_{--} & B_{-+} & 0\\
0 & B_{+-} & B_{++} & 0\\
0 & 0 & 0 & 1
\end{array}
\right).\label{dAd}
\ee
If $B=u\in SU(2)$ one obtains 
\be
e^{-(b^{\dag}Ab+d^{\dag}Ad)}b_se^{b^{\dag}Ab+d^{\dag}Ad}
=
\sum_{s'}u_{ss'}b_{s'},\quad
e^{-(b^{\dag}Ab+d^{\dag}Ad)}d_se^{b^{\dag}Ab+d^{\dag}Ad}
=
\sum_{s'}u_{ss'}d_{s'}.\label{ud}
\ee
Similarly one can show that 
\be
e^{-i(\alpha b^{\dag}b+\beta d^{\dag}d)}b_se^{i(\alpha b^{\dag}b+\beta d^{\dag}d)}
=
e^{i\alpha}b_{s},\quad
e^{-i(\alpha b^{\dag}b+\beta d^{\dag}d)}d_se^{i(\alpha b^{\dag}b+\beta d^{\dag}d)}
=
e^{i\beta}d_{s}.\label{Qd}
\ee
Of particular interest are the important formulas 
\be
e^{-(b^{\dag}Ab+d^{\dag}Ad)}\bbox I_0 e^{b^{\dag}Ab+d^{\dag}Ad}
=
\bbox I_0,\quad
e^{-i(\alpha b^{\dag}b+\beta d^{\dag}d)}\bbox I_0
e^{i(\alpha b^{\dag}b+\beta d^{\dag}d)}
=
\bbox I_0.\label{QI}
\ee
The vacuum of the representation is
\be
|0,0,0,0\rangle=
\left(\begin{array}{c}
0\\
1
\end{array}
\right)\otimes 
\left(\begin{array}{c}
0\\
1
\end{array}
\right)\otimes 
\left(\begin{array}{c}
0\\
1
\end{array}
\right)\otimes 
\left(\begin{array}{c}
0\\
1
\end{array}
\right)\label{0000}
\ee
and satisfies
\be
\bbox I_0|0,0,0,0\rangle
=
|0,0,0,0\rangle,\quad
e^{b^{\dag}Ab+d^{\dag}Ad}|0,0,0,0\rangle
=
|0,0,0,0\rangle.
\ee
Creation operators act on vacuum by 
\be
b_-^{\dag}|0,0,0,0\rangle
=
|1,0,0,0\rangle,\quad
b_+^{\dag}|0,0,0,0\rangle
=
|0,1,0,0\rangle,\quad
d_-^{\dag}|0,0,0,0\rangle
=
|0,0,1,0\rangle,\quad
d_+^{\dag}|0,0,0,0\rangle
=
|0,0,0,1\rangle.
\ee
The above representation of CAR is a starting point for the Poincar\'e covariant reducible representation of CAR we shall discuss in the next sections. 

\section{Poincar\'e transformations of free Dirac fields}

The Dirac equation for a free electron
can be written in the 2-spinor form \cite{PR}
as 
\be
\left(
\begin{array}{cc}
\frac{m}{\sqrt{2}}
\varepsilon{_A}{^B} &
-i\nabla{_{A}}{^{B'}}\\
i\nabla{^{B}}{_{A'}} &
\frac{m}{\sqrt{2}}
\varepsilon{_{A'}}{^{B'}}
\end{array}
\right)
\left(
\begin{array}{c}
\psi_B\\
\psi_{B'}
\end{array}
\right)=0.
\ee
We use the units with $\hbar=1$ and $c=1$. The
energy-momentum world-vector $p^a=(\sqrt{\bbox p^2+m^2},\bbox p)$ can
be decomposed in terms of two null directions \cite{BW}
$
p^a=\pi^{a}+\frac{m^2}{2}\omega^{a}
=
\pi^{A} \bar \pi^{A'}
+
\frac{m^2}{2}
\omega^{A} \bar \omega^{A'}
,\label{p}
$
satisfying the spin-frame condition $\omega_{A}\pi^A=1$.

The 2-spinor language naturally privileges null directions \cite{PR}. 
The projections $S(\omega,p)=\omega^aS_a$ 
of the Pauli-Lubanski vector in the  $\omega^A\bar \omega^{A'}$ direction are \cite{BW}
\be
S(\omega,p){_{A}}{^{B}}=
\frac{1}{2}\Bigl(
\pi_{A}\omega^{B}+ \omega_{A}\pi^{B}\Bigr),\quad
S(\omega,p){_{A'}}{^{B'}}=-
\frac{1}{2}\Bigl(
\bar \pi_{A'}\bar \omega^{B'}+ 
\bar \omega_{A'}\bar \pi^{B'}\Bigr)\label{PL2}.
\ee
The eigenvectors 
\be
{\phi}_{\pm,\alpha}^{(+)}(\bbox p)=
\left(
\begin{array}{c}
\pm\frac{m}{\sqrt{2}}
\omega{_{A}}\\
- \bar \pi{_{A'}}
\end{array}
\right),\quad
{\phi}_{\pm,\alpha}^{(-)}(\bbox p)=
\left(
\begin{array}{c}
- \pi{_{A}}\\
\mp\frac{m}{\sqrt{2}}
\bar \omega{_{A'}}
\end{array}
\right),\label{I.41}
\ee 
correspond to the four combinations of signs of spin and frequency. 
The subscripts $\pm$ denote the signs of frequency and the superscripts
$(\pm)$ the signs of spin.

The invariant measure on the mass-$m$ hyperboloid is 
$
d\Gamma_m(\bbox p)
=
\frac{1}{(2\pi)^3}\frac{d^3p}{2\sqrt{\bbox p^2+m^2}}.
$
A general solution of the free Dirac equation can be written in a
Fourier form as
\be
\psi_\alpha(x)
&=&
\sum_s\int d \Gamma_m(\bbox p)\Big(
{\phi}_{+,\alpha}^{(s)}(\bbox p) f(\bbox p,s)e^{-ip\cdot x}
+
{\phi}_{-,\alpha}^{(s)}(\bbox p) \overline{g(\bbox p,-s)}e^{ip\cdot x}
\Big)
\ee
The charge conjugated solution is
\be
\psi^c_\alpha(x)
&=&
\sum_s\int d \Gamma_m(\bbox p)\Big(
{\phi}_{+,\alpha}^{(s)}(\bbox p) g(\bbox p,s)e^{-ip\cdot x}
+
{\phi}_{-,\alpha}^{(s)}(\bbox p) \overline{f(\bbox p,-s)}e^{ip\cdot x}
\Big).
\ee
The Dirac equation is associated with two representations of (the
covering space of) the Poincar\'e group which are dual
to each other (the active finite dimensional nonunitary spinor
representation and
the passive infinite dimensional unitary representation).

Denote by
$\Lambda_a{^b}$, $\Lambda_{\alpha}{^{\beta}}$, 
$\Lambda_{A}{^{B}}$, and $\Lambda_{A'}{^{B'}}$, respectively, the
representations $(1/2,1/2)$, $(1/2,0)\oplus(0,1/2)$, 
$(1/2,0)$,  and $(0,1/2)$ of $\Lambda\in SL(2,C)$. 
The active transformation of the bispinor field 
$
T_{S,y}\psi_\alpha(x)
=
\Lambda_{\alpha}{^{\beta}}\psi_\beta\big(\Lambda^{-1}(x-y)\big)
$
induces the passive transformation of the amplitudes
by means of 
\be
T_{\Lambda,y}\psi_\alpha(x)
&=&
\sum_s\int d \Gamma_m(\bbox p)\Big(
{\phi}_{+,\alpha}^{(s)}(\bbox p) 
{\cal T}_{\Lambda,y}f(\bbox p,s)e^{-ip\cdot x}
+
{\phi}_{-,\alpha}^{(s)}(\bbox p) 
\overline{{\cal T}_{\Lambda,y}g(\bbox p,-s)}e^{ip\cdot x}
\Big),\\
\left(
\begin{array}{c}
{\cal T}_{\Lambda,y}f(\bbox p,-)\\
{\cal T}_{\Lambda,y}f(\bbox p,+)
\end{array}
\right)
&=&
e^{ip\cdot y}
\underbrace{
\left(
\begin{array}{cc}
\omega{_{A}}(\bbox p)\Lambda\pi{^{A}}(\bbox p) & 
-\frac{m}{\sqrt{2}}\omega{_{A}}(\bbox p)
\Lambda\omega{^{A}}(\bbox p)\\
\frac{m}{\sqrt{2}}
{\bar \omega}{_{A'}}(\bbox p)\overline{\Lambda\omega}{^{A'}}(\bbox p) &
{\bar \omega}{_{A'}}(\bbox p)\overline{\Lambda\pi}{^{A'}}(\bbox p)
\end{array}
\right)}_{u(\Lambda,\bbox p)\in SU(2)}
\left(
\begin{array}{c}
f(\bbox{\Lambda^{-1}p},-)\\
f(\bbox{\Lambda^{-1}p},+)
\end{array}
\right).\label{matrix}
\ee
The amplitudes $g$ transform in the same way as $f$. The transformed spin-frames are
$
\Lambda\omega{_{A}}(\bbox p)
=
\Lambda{_A}{^B}\omega{_{B}}(\bbox{\Lambda^{-1}p})
$,
$
\Lambda\pi{_{A}}(\bbox p)
=
\Lambda{_A}{^B}\pi{_{B}}(\bbox{\Lambda^{-1}p})$.
The representation (\ref{matrix}) is unitary with respect to
the positive-definite scalar product
\be
\langle f_1|f_2\rangle=
\sum_s\int d \Gamma_m(\bbox p)\overline{f_1(\bbox p,s)}f_2(\bbox p,s).
\ee	
\section{Reducible quantization}

We follow the strategy proposed in \cite{1,2}. 
Take four CAR operators $b_\pm$,  $d_\pm$ we have discussed above and introduce the
following four operators 
\be
b(\bbox p,s)
=
|\bbox p\rangle\langle \bbox p|\otimes b_s=c_1(\bbox p,s),\quad
d(\bbox p,s)
=
|\bbox p\rangle\langle \bbox p|\otimes d_s=c_2(\bbox p,s).
\ee
The momentum eigenvectors are normalized by 
$
\langle \bbox p|\bbox p'\rangle
=
\delta_{\Gamma_m}(\bbox p,\bbox p')
=
(2\pi)^3 2\sqrt{\bbox p^2+m^2}\delta^{(3)}(\bbox p-\bbox p').
$
The reducible representation of CAR can be written in a compact form as 
\be
\big\{c_n(\bbox p,s),c_{n'}(\bbox p',s')^{\dag}\big\}
=
\delta_{nn'}\delta_{ss'}\delta_{\Gamma_m}(\bbox p,\bbox p')
|\bbox p\rangle\langle \bbox p|\otimes \bbox I
=
\delta_{nn'}\delta_{ss'}\delta_{\Gamma_m}(\bbox p,\bbox p')
I_{\bbox p},\label{nCAR}
\ee
the remaining anti-commutators vanishing.
The identity at the right side of (\ref{nCAR}) is the one
occuring in the CAR relations (\ref{CAR})
and the RHS of (\ref{nCAR}) is in the center of the CAR algebra,
i.e. commutes with all CAR operators. 
Similarly to the CCR case \cite{1,2,3} we 
have introduced the operator 
$
I_{\bbox p}=|\bbox p\rangle\langle \bbox p|\otimes \bbox I
$
satisfying the resolution of unity 
$
\int d \Gamma_m(\bbox p)
I_{\bbox p}=I.
$
We define the single-oscillator Dirac field operator by 
\be
\Psi_\alpha(x)=\sum_s\int d \Gamma_m(\bbox p)\Big(
{\phi}_{+,\alpha}^{(s)}(\bbox p) b(\bbox p,s)e^{-ip\cdot x}
+
{\phi}_{-,\alpha}^{(s)}(\bbox p) d(\bbox p,-s)^{\dag}e^{ip\cdot x}
\Big).
\ee
The charge conjugated operator reads
\be
\Psi^c_\alpha(x)=\sum_s\int d \Gamma_m(\bbox p)\Big(
{\phi}_{+,\alpha}^{(s)}(\bbox p) d(\bbox p,s)e^{-ip\cdot x}
+
{\phi}_{-,\alpha}^{(s)}(\bbox p) b(\bbox p,-s)^{\dag}e^{ip\cdot x}
\Big).
\ee
In order to perform the second step of quantization we introduce 
\be
I_0=\int d \Gamma_m(\bbox p)|\bbox p\rangle\langle \bbox p|\otimes \bbox I_0.
\ee
The $N$-oscillator extension is defined by 
\be
\uu b(\bbox p,s)
&=&
\frac{1}{\sqrt{N}}\Big(
b(\bbox p,s)\otimes I\otimes\dots\otimes I
+
I_0\otimes b(\bbox p,s)\otimes I \otimes\dots\otimes I
+
\dots
+
I_0\otimes\dots\otimes I_0 \otimes b(\bbox p,s)
\Big)
=
\uu c_1(\bbox p,s),\\
\uu d(\bbox p,s)
&=&
\frac{1}{\sqrt{N}}\Big(
d(\bbox p,s)\otimes I\otimes\dots\otimes I
+
I_0\otimes d(\bbox p,s)\otimes I \otimes\dots\otimes I
+
\dots
+
I_0\otimes\dots\otimes I_0 \otimes d(\bbox p,s)
\Big)
=
\uu c_2(\bbox p,s),\\
\uu\Psi_\alpha(x)
&=&
\sum_s\int d \Gamma_m(\bbox p)\Big(
{\phi}_{+,\alpha}^{(s)}(\bbox p) \uu b(\bbox p,s)e^{-ip\cdot x}
+
{\phi}_{-,\alpha}^{(s)}(\bbox p) \uu d(\bbox p,-s)^{\dag}e^{ip\cdot x}
\Big),\\
\uu \Psi^c_\alpha(x)
&=&
\sum_s\int d \Gamma_m(\bbox p)\Big(
{\phi}_{+,\alpha}^{(s)}(\bbox p) \uu d(\bbox p,s)e^{-ip\cdot x}
+
{\phi}_{-,\alpha}^{(s)}(\bbox p) \uu b(\bbox p,-s)^{\dag}e^{ip\cdot x}
\Big).
\ee
The reducible representation of CAR reads
\be
\big\{\uu c_n(\bbox p,s),\uu c_{n'}(\bbox p',s')^{\dag}\big\}
&=&
\delta_{nn'}\delta_{ss'}\delta_{\Gamma_m}(\bbox p,\bbox p')
\uu I_{\bbox p},\label{uu nCAR}
\ee
with 
$
\uu I_{\bbox p}
=
\frac{1}{N}
\Big(
I_{\bbox p}\otimes I\otimes \dots \otimes I
+
\dots
+
I\otimes \dots \otimes I\otimes I_{\bbox p}
\Big).
$
\section{Field operators are indeed operators}

$\uu\Psi_\alpha(x)$ has been defined as an $N$-oscillator extension of a 
single-oscillator $\Psi_\alpha(x)$. Therefore, in order to verify that $\uu\Psi_\alpha(x)$ is an operator it is sufficient to check this property for $\Psi_\alpha(x)$. 
The choice of the representation implies that 
\be
\Psi_\alpha(x)
=
\sum_s\Big(
{\phi}_{+,\alpha}^{(s)}(\hat{\bbox p}) W(x)\otimes b_s
+
{\phi}_{-,\alpha}^{(s)}(\hat{\bbox p}) W(x)^{\dag} \otimes d_{-s}^{\dag}
\Big)\nonumber
\ee
where 
$
\hat{\bbox p}=\sum_s\int d \Gamma_m(\bbox p)
\bbox p|\bbox p\rangle\langle \bbox p|
$
is the spectral representation of an unbounded operator, 
$
{\phi}_{\pm,\alpha}^{(s)}(\hat{\bbox p})
$
are functions of the operator $\hat{\bbox p}$ in the sense of spectral theory, and
$
W(x)=\int d \Gamma_m(\bbox p)
|\bbox p\rangle\langle \bbox p|e^{-ip\cdot x}
$
is unitary. All these objects are well defined and there is no problem with products of fields taken at the same point $x$ of the configuration space. 
The difference between fields taken in our reducible representation and those arising from the standard Fock construction is analogous to this between the unitary operator $W(x)$ and the distribution $\int d \Gamma_m(\bbox p)
e^{-ip\cdot x}$.

\section{Vacuum and multi-electron states}

The vacuum consists of a Hilbert space of all the states which are annihilated by all annihilation operators. We begin with a ``single-oscillator vacuum" 
\be
|O\rangle
&=&
\sum_s\int d \Gamma_m(\bbox p)O(\bbox p)|\bbox p,0,0,0,0\rangle,
\ee
where 
$
|\bbox p,0,0,0,0\rangle
=
|\bbox p\rangle|0,0,0,0\rangle
$
and $|0,0,0,0\rangle$ is defined by (\ref{0000}). One finds indeed 
$
b(\bbox p,s)|O\rangle=d(\bbox p,s)|O\rangle=0.
$
The vacuum is defined at the $N$-oscillator level as the tensor product of one-oscilator vacua
\be
|\uu O\rangle
=
\underbrace{|O\rangle\dots|O\rangle}_N.
\ee
As expected
$
\uu b(\bbox p,s)|\uu O\rangle=\uu d(\bbox p,s)|\uu O\rangle=0.
$

To discuss multi-electron states it is convenient to introduce the
smeared out CAR operators
\be
\uu c_n(f)
=
\sum_{s}\int d \Gamma_m(\bbox p) \overline{f(\bbox p,s)}
\uu c_n(\bbox p,s),\quad
\uu c_n(f)^{\dag}
=
\sum_{s}\int d \Gamma_m(\bbox p) f(\bbox p,s)\uu c_n(\bbox p,s)^{\dag},
\ee
satisfying
\be
\big\{\uu c_n(f),\uu c_{n'}(g)^{\dag}\big\}
&=&
\delta_{nn'}\sum_{s}\int d \Gamma_m(\bbox p) \overline{f(\bbox p,s)}
g(\bbox p,s)\uu I_{\bbox p}. \label{nonCAR'}
\ee
The RHS of (\ref{nonCAR'}) is in the center of the CAR algebra. 

The scalar product of two unnormalized one-electron states is 
\be
\langle \uu O|\uu c_n(f) \uu c_{n'}(g)^{\dag}|\uu O\rangle
=
\delta_{nn'}\sum_{s}\int d \Gamma_m(\bbox p)Z(\bbox p)
\overline{f(\bbox p,s)}g(\bbox p,s)
=
\delta_{nn'}\langle Of|Og\rangle=:
\delta_{nn'}\langle f|g\rangle_Z,\label{<of|og>}
\ee
with $Z(\bbox p)=|O(\bbox p)|^2$. 
The scalar product at (\ref{<of|og>}) is analogous to the formula for
one-photon states given in \cite{3}. 
$fO$ denotes the pointlike product 
$fO(\bbox p,s)=O(\bbox p)f(\bbox p,s)$. 

The
following important result can be proved in a direct analogy to Theorem
1 from \cite{3} so we leave the proof to the reader. Denote by $\sum_\sigma$ the sum over all the
permutations of the set $\{1,\dots,M\}$.
\medskip

\noindent
{\bf Theorem 1.} 
\be
{}&{}&\lim_{N\to\infty}
\langle \uu O|\uu c_n(f_1)\dots \uu c_n(f_M)\uu c_n(g_1)^{\dag}\dots
\uu c_n(g_M)^{\dag}|\uu O\rangle
=
\sum_{\sigma}\delta_\sigma
\langle f_1|g_{\sigma(1)}\rangle_Z
\dots 
\langle f_M|g_{\sigma(M)}\rangle_Z\nonumber\\
&{}&\pp{==}=
\sum_{\sigma}\delta_\sigma\sum_{s_1\dots s_M}
\int d\Gamma_m(\bbox p_1)Z(\bbox p_1)\dots
d\Gamma_m(\bbox p_M)Z(\bbox p_M)
\overline{f_1(\bbox p_1,s_1)}\dots 
\overline{f_m(\bbox p_M,s_M)}
g_{\sigma(1)}(\bbox p_1,s_1)
\dots 
g_{\sigma(m)}(\bbox p_M,s_M)\nonumber
\ee
where $\delta_\sigma$ is the sign of the permutation $\sigma$.
\medskip
\rule{5pt}{5pt}

\section{Action of the Poincar\'e group on field operators}

For any $\Lambda\in SL(2,\bbox C)$, $y\in \bbox T^{(1,3)}$ we will construct a representation of (the covering of) the Poincar\'e group, 
$
(\Lambda,y)\mapsto
\uu U_{\Lambda,y}$, acting by 
$
\uu U_{\Lambda,y}^{\dag}
\uu \Psi_{\alpha}(x)
\uu U_{\Lambda,y}
=
\Lambda{_{\alpha}}{^\beta}\uu \Psi_{\beta}\big(\Lambda^{-1}(x-y)\big).
$
The map $\Lambda\mapsto \Lambda{_{\alpha}}{^\beta}$
is the bispinor representation of $\Lambda$. We will begin with constructing
$
U_{\Lambda,y}^{\dag}
\Psi_{\alpha}(x)
U_{\Lambda,y}
=
\Lambda{_{\alpha}}{^\beta}\Psi_{\beta}\big(\Lambda^{-1}(x-y)\big)
$,
which additionally satisfies 
$
U_{\Lambda,y}^{\dag}
I_0
U_{\Lambda,y}
=
I_0.
$
Once we have completed this stage the final representation is 
$
\uu U_{\Lambda,y}
=
\underbrace{U_{\Lambda,y}\otimes\dots\otimes U_{\Lambda,y}}_N.
$

The generator of $U_{\bbox 1,y}$ is chosen in a way which is consistent with an appropriate ``single-oscillator" Noether invariant, i.e.
\be
P_a
=
\int d\Gamma_m(\bbox p)p_a|\bbox p\rangle\langle \bbox p|\otimes
\frac{1}{2}
\big(b^{\dag}b-bb^{\dag}
+
d^{\dag}d-dd^{\dag}
\big)
=
\int d\Gamma_m(\bbox p)p_a|\bbox p\rangle\langle \bbox p|\otimes
\big(b^{\dag}b+d^{\dag}d-2\bbox I\big).
\ee
The part $-2\int d\Gamma_m(\bbox p)p_a|\bbox p\rangle\langle \bbox p|\otimes \bbox I$ belongs to the center of CAR. 
Then, using the results from Sec.~I, we can show that 
$
U_{\bbox 1,y}
=
e^{iy\cdot P}
$
acts by
\be
U_{\bbox 1,y}^{\dag}
b(\bbox p,s)
U_{\bbox 1,y}
&=&
|\bbox p\rangle\langle \bbox p|\otimes 
e^{-iy\cdot p(b^{\dag}b+d^{\dag}d)}b_se^{iy\cdot p(b^{\dag}b+d^{\dag}d)}
=
e^{iy\cdot p}b(\bbox p,s)\\
U_{\bbox 1,y}^{\dag}
d(\bbox p,s)
U_{\bbox 1,y}
&=&
|\bbox p\rangle\langle \bbox p|\otimes 
e^{-iy\cdot p(b^{\dag}b+d^{\dag}d)}d_se^{iy\cdot p(b^{\dag}b+d^{\dag}d)}
=
e^{iy\cdot p}d(\bbox p,s),\\
U_{\bbox 1,y}^{\dag}
I_0
U_{\bbox 1,y}
&=&
\int d\Gamma_m(\bbox p)|\bbox p\rangle\langle \bbox p|\otimes
e^{-iy\cdot p(b^{\dag}b+d^{\dag}d)}\bbox I_0e^{iy\cdot p(b^{\dag}b+d^{\dag}d)}
=
I_0.
\ee
Finally, 
$
\uu U_{\bbox 1,y}^{\dag}
\uu \Psi_{\alpha}(x)
\uu U_{\bbox 1,y}
=
\uu \Psi_{\alpha}(x-y).
$
The four-momentum at the $N$-oscillator level reads 
\be
\uu P_a
=
P_a\otimes I\otimes \dots\otimes I
+
\dots
+
I\otimes \dots\otimes I\otimes P_a
\ee
and satisfies 
\be
{[\uu P_a,\uu c_n(\bbox p,s)^{\dag}]}
&=&
p_a\uu c_n(\bbox p,s)^{\dag}
\ee
which is the same result as in the Fock representation. 

Now consider 
\be
U_{\Lambda,0} 
&=&
\exp\Big(i\int d\Gamma_m(\bbox p)|\bbox p\rangle \langle \bbox p|
\otimes \big(b^{\dag}A(\Lambda,\bbox p)b+d^{\dag}A(\Lambda,\bbox p)d\big)\Big)
\Big(\int d\Gamma_m(\bbox p')|\bbox p'\rangle
\langle \bbox {\Lambda^{-1}p'}|\otimes  \bbox I\Big)\\
&=&
\int d\Gamma_m(\bbox p)|\bbox p\rangle \langle \bbox p|
\otimes \exp\Big(i\big(b^{\dag}A(\Lambda,\bbox p)b+d^{\dag}A(\Lambda,\bbox p)d\big)\Big)
\Big(\int d\Gamma_m(\bbox p')|\bbox p'\rangle
\langle \bbox {\Lambda^{-1}p'}|\otimes  \bbox I\Big)
\ee
where $u(\Lambda,\bbox p)=e^{iA(\Lambda,\bbox p)}\in SU(2)$ is the matrix occuring in (\ref{matrix}). An equivalent definition employing directly 
$u(\Lambda,\bbox p)$, instead of $A(\Lambda,\bbox p)$, can be formulated by means of (\ref{dAd}). Taking into account (\ref{ud}) one finds 
\be
\left(
\begin{array}{c}
U_{\Lambda,0}^{\dag}c_n(\bbox p,-)U_{\Lambda,0}\\
U_{\Lambda,0}^{\dag}c_n(\bbox p,+)U_{\Lambda,0}
\end{array}
\right)
&=&
\left(
\begin{array}{cc}
\omega{_{A}}(\bbox p)\Lambda\pi{^{A}}(\bbox p) & 
-\frac{m}{\sqrt{2}}\omega{_{A}}(\bbox p)
\Lambda\omega{^{A}}(\bbox p)\\
\frac{m}{\sqrt{2}}
{\bar \omega}{_{A'}}(\bbox p)\overline{\Lambda\omega}{^{A'}}(\bbox p) &
{\bar \omega}{_{A'}}(\bbox p)\overline{\Lambda\pi}{^{A'}}(\bbox p)
\end{array}
\right)
\left(
\begin{array}{c}
c_n(\bbox{\Lambda^{-1}p},-)\\
c_n(\bbox{\Lambda^{-1}p},+)
\end{array}
\right),\quad
U_{\Lambda,0}^{\dag}I_0 U_{\Lambda,0}
=
I_0
\ee
and finally 
\be
\left(
\begin{array}{c}
\uu U_{\Lambda,0}^{\dag}\uu c_n(\bbox p,-)\uu U_{\Lambda,0}\\
\uu U_{\Lambda,0}^{\dag}\uu c_n(\bbox p,+)\uu U_{\Lambda,0}
\end{array}
\right)
&=&
\left(
\begin{array}{cc}
\omega{_{A}}(\bbox p)\Lambda\pi{^{A}}(\bbox p) & 
-\frac{m}{\sqrt{2}}\omega{_{A}}(\bbox p)
\Lambda\omega{^{A}}(\bbox p)\\
\frac{m}{\sqrt{2}}
{\bar \omega}{_{A'}}(\bbox p)\overline{\Lambda\omega}{^{A'}}(\bbox p) &
{\bar \omega}{_{A'}}(\bbox p)\overline{\Lambda\pi}{^{A'}}(\bbox p)
\end{array}
\right)
\left(
\begin{array}{c}
\uu c_n(\bbox{\Lambda^{-1}p},-)\\
\uu c_n(\bbox{\Lambda^{-1}p},+)
\end{array}
\right),\\
\uu U_{\Lambda,0}^{\dag}
\uu \Psi_{\alpha}(x)
\uu U_{\Lambda,0}
&=&
\Lambda{_{\alpha}}{^\beta}\uu \Psi_{\beta}(\Lambda^{-1}x).
\ee

\section{Action of the Poincar\'e group on states}

It is sufficient to concentrate on the covariance properties of vacuum; the 
multi-electron states transform according to the properties of vacuum and those of the creation operators. 
Since 
\be
\uu U_{\Lambda,y}|\uu O\rangle
=
U_{\Lambda,y}|O\rangle\otimes\dots \otimes U_{\Lambda,y}|O\rangle
\ee
we explicitly write only 
\be
U_{\Lambda,y}|O\rangle
&=&
\int d\Gamma_m(\bbox p)e^{-2iy\cdot p}O(\bbox {\Lambda^{-1}p})|\bbox p,0,0,0,0\rangle.
\ee
As we can see the vacua are not invariant but only covariant. The factor 
$e^{-2iy\cdot p}$ can be removed by a unitary transformation which belongs to the center of CAR. In this new representation the four momentum annihilates vacuum. 
One should stress that the choice of the ``vacuum representation" is here well defined (as opposed to the standard Fock prescription, where the phase is infinite). The possibility of such a choice of representation does not physically mean that the vacuum energy is zero. In the vacuum representation the vacuum transforms by 
\be
O(\bbox p)\mapsto
O(\bbox {\Lambda^{-1}p})=:V_{\Lambda,y}O(\bbox p)
\ee
and 
$
\uu U_{\Lambda,y}^{\dag}\uu c_n(\bbox p,s)\uu U_{\Lambda,y}
=
\uu V_{\Lambda,y}^{\dag}\uu c_n(\bbox p,s)\uu V_{\Lambda,y}.
$
\section{Charge}

The operator 
\be
Q
&=&
e_0 \int d \Gamma_m(\bbox p)|\bbox p\rangle\langle \bbox p|\otimes
(b^{\dag}b+dd^{\dag})
=
e_0 \int d \Gamma_m(\bbox p)|\bbox p\rangle\langle \bbox p|\otimes
(b^{\dag}b-d^{\dag}d+2\bbox I)
\ee
is the Noether invariant corresponding to the $U(1)$ symmetry of the free Dirac Lagrangian at a single-oscillator level. Formulas (\ref{Qd}), (\ref{QI}), imply 
\be
e^{-i\varphi Q}\Psi_\alpha(x)e^{i\varphi Q}
=
e^{-ie_0\varphi}\Psi_\alpha(x),\quad
e^{-i\varphi Q}\Psi^c_\alpha(x)e^{i\varphi Q}
=
e^{+ie_0\varphi}\Psi^c_\alpha(x),\quad
e^{-i\varphi Q}\bbox I_0e^{i\varphi Q}
=
\bbox I_0.
\ee
Accordingly 
$
e^{i\varphi \uu Q}
=
e^{i\varphi Q}\otimes\dots\otimes e^{i\varphi Q}
$
is the $U(1)$ gauge transformation 
\be
e^{-i\varphi \uu Q}\uu \Psi_\alpha(x)e^{i\varphi \uu Q}
=
e^{-ie_0\varphi}\uu \Psi_\alpha(x),\quad
e^{-i\varphi \uu Q}\uu \Psi^c_\alpha(x)e^{i\varphi \uu Q}
=
e^{+ie_0\varphi}\uu \Psi^c_\alpha(x).
\ee
The generator 
$
\uu Q
=
Q\otimes I\otimes \dots\otimes I
+
\dots
+
I\otimes \dots\otimes I\otimes Q
$
plays the role of the charge operator. 

Charge of multi-electron states follows from 
\be
{[\uu Q,\uu b(\bbox p,s)^{\dag}]}
=
+e_0\uu b(\bbox p,s)^{\dag},\quad
{[\uu Q,\uu d(\bbox p,s)^{\dag}]}
=
-e_0\uu d(\bbox p,s)^{\dag}
\ee
which is the same result as in the Fock representation. 

\section{Spin}

The generators of the unitary representation (\ref{matrix}) of
$SL(2,C)$ are
$
j^{ab}=l^{ab}+s^{ab}
$
where $l^{ab}$ is the usual orbital part in momentum representation
and 
\be
s^{ab}
&=&
\left(
\begin{array}{cc}
\omega{_{X}}(p)J^{ab}{^{X}}{^{Y}}\pi{_{Y}}(p) & 
-\frac{m}{\sqrt{2}}\omega{_{X}}(p)
J^{ab}{^{X}}{^{Y}}\omega{_{Y}}(p)\\
\frac{m}{\sqrt{2}}
{\bar \omega}{_{X'}}(p)J^{ab}{^{X'}}{^{Y'}}\bar \omega{_{Y'}}(p) &
{\bar \omega}{_{X'}}(p)J^{ab}{^{X'}}{^{Y'}}\bar \pi{_{Y'}}(p)
\end{array}
\right).\label{s^ab}
\ee
The operators
\be
J^{ab}{_{X}}{^{Y}}
=
L^{ab}\ve{_{X}}{^{Y}}
+
\frac{i}{2}\ve{^{A'}}{^{B'}}
\Bigl(
\ve{^{A}}{_{X}}\ve{^{B}}{^{Y}}
+
\ve{^{B}}{_{X}}\ve{^{A}}{^{Y}}
\Bigr),\quad
J^{ab}{_{X'}}{^{Y'}}
=
L^{ab}\ve{_{X'}}{^{Y'}}
+
\frac{i}{2}\ve{^{A}}{^{B}}
\Bigl(
\ve{^{A'}}{_{X'}}\ve{^{B'}}{^{Y'}}
+
\ve{^{B'}}{_{X'}}\ve{^{A'}}{^{Y'}}
\Bigr),
\ee
generate spinor $(1/2,0)$ and $(0,1/2)$ representations. Hermiticity
of $s^{ab}$ is not explicit but can be proved by means of certain
spinor identities valid for spin-frames \cite{BW}. Let us note that
the matrix $s^{ab}$ involves also the matrix elements of $l^{ab}$
evaluated between the spin-frame spinors. This ``orbital" element
occuring in the ``spin" part of the generators is a consequence of
the fact that spin-frames occuring in the $SU(2)$ matrix are
spinor fields themselves. 

It follows that the generators of $U_{\Lambda,0}$ are 
$
J^{ab}=L^{ab}+S^{ab}
$
where $L^{ab}$ is the orbital part and 
\be
S^{ab}(\hat {\bbox p})
&=&
\int d\Gamma_m(\bbox p)|\bbox p\rangle\langle \bbox p|
\otimes
\Big(
b^{\dag}s^{ab}(\bbox p)b+
d^{\dag}s^{ab}(\bbox p) d
\Big).
\ee
If $p^a$ is future-pointing and $\omega^a$ is the associated null
vector then the projection of the Pauli-Lubanski vector $s^a$ in the direction of
$\omega^a$, corresponding to the representation (\ref{matrix}) and
(\ref{s^ab}), reads
\be
s=\omega^a(\bbox p) s_a(\bbox p)
=
\left(
\begin{array}{cc}
-\frac{1}{2} & 0\\
0 & \frac{1}{2}
\end{array}
\right)=-\frac{1}{2}\sigma_3.
\ee
We define the spin operator for $U_{\Lambda,y}$ in an analogous way by 
$
S^a(\hat {\bbox p})
=
\hat p_b {^*}S^{ab}(\hat {\bbox p}).
$
Let us note that we use here 
$
\hat p_b
=
\int d\Gamma_m(\bbox p)p_b|\bbox p\rangle\langle \bbox p|
$
and not 
$
P_b
=
\int d\Gamma_m(\bbox p)p_b|\bbox p\rangle\langle \bbox p|\otimes
\big(b^{\dag}b+d^{\dag}d-2\bbox I\big).
$
The projection 
\be
S&=&\omega^a(\hat{\bbox p}) S_a(\hat{\bbox p})
=
\int d\Gamma_m(\bbox p)|\bbox p\rangle\langle \bbox p|
\otimes
\Big(
b^{\dag}sb+
d^{\dag} s d
\Big)
=
\frac{1}{2}
\int d\Gamma_m(\bbox p)|\bbox p\rangle\langle \bbox p|
\otimes
\Big(
b_+^{\dag}b_+
-
b_-^{\dag}b_-
+
d_+^{\dag}d_+
-
d_-^{\dag}d_-
\Big)
\ee
plays a role analogous to the helicity operator. 

At the $N$-oscillator level the spin operator is defined in terms of generators of $\uu U_{\Lambda,0}$ and reads 
\be
\uu S
&=&
S\otimes I\otimes \dots\otimes I
+
\dots
+
I\otimes \dots\otimes I\otimes S.
\ee
One can verify that spin of vacuum is zero
$
\uu S|\uu O\rangle=0
$
which is consistent with the fact that vacuum transforms as a scalar field. 
Spin of multi-electron states follows from 
\be
{[\uu S,\uu c_n(\bbox p,s)^{\dag}]}=s\uu c_n(\bbox p,s)^{\dag}
\ee
which is the same result as in the Fock representation. 

\section{Discussion}

Theorem 1 as well as the analogous result from \cite{3} suggest that perturbative expansions associated with the Dyson series will have the same form (for $N\to\infty$) in the reducible framework as they have in the Fock one, with one modification: The scalar products $\langle\cdot|\cdot\rangle$ will be replaced by $\langle\cdot|\cdot\rangle_Z$. The functions $Z(\bbox k)$
(for photons) and $Z(\bbox p)$ (for electrons) cannot be identified since they transform as scalar fields with zero and non-zero masses, respectively. As discussed in \cite{1,2,3} not only do these functions  regularize integrals but they appear, for $N\to\infty$, in exactly those places where one expects renormalization constants $Z_3$ and $Z_2$. Naudts et al. \cite{NKdR} have recently proposed to associate $Z(\bbox k)$, $Z(\bbox p)$ with fields of space-time fluctuations. Assuming that the functions are constant up to the Planck energies, or perhaps some other scale \cite{Saini}, 
the only modification one expects within the quantum electrodynamics regime is in a rescaling of bare parameters into observable ones. 

Interesting modifications occur if one considers vacuum energies. Let us note that the Noether invariants \cite{Noether} did not need to be normally ordered, since the vacuum contributions were well behaved. For example, the energy of 
$N_B$ bosonic and $N_F$ fermionic oscillators described by the vacua 
$Z_B(\bbox k)$, $Z_F(\bbox p)$ is 
\be
\langle E\rangle_{\rm vac}
&=&
N_B \int d\Gamma_0(\bbox k)|\bbox k| Z_B(\bbox k)
-
2 N_F \int d\Gamma_m(\bbox p)\sqrt{\bbox p^2+m^2}
Z_F(\bbox p). 
\ee
The expression becomes infinite only in the limits of infinite numbers of oscillators. It is essential, however, that the limits we have used, say, in Theorem 1 are understood in the sense of thermodynamic limits of ``very large numbers". 
The vacuum energy can be thus positive, negative,
or zero, depending on the vacuum states. In any case it can be finite 
if the vacua are appropriately chosen. Such a regularization of the
vacuum divergence may have important implications for the cosmological
constant problem. 

Another important modification with respect to the standard Fock-type quantum field theory is that the reducible quantization introduces a kind of nonlocality whose scale is given, effectively, by the shapes of vacuum wave functions. For this reason the formalism we have obtained does not fit into the framework of generalized free fields \cite{Greenberg,BLT} where the right-hand-sides of CCR/CAR involve functions of Casimirs of symmetry groups. The reducible quantization  approach is closer to nonlocal quantum field theories 
\cite{Efimov,Moffat,Evens,Kleppe,Clayton,Cornish,Basu}. 

The main open question is to what extent our formalism is applicable to concrete calculations in perturbative quantum electrodynamics, and whether all the infinities are indeed automatically regularized to all orders of perturbation theory. It seems that various techniques  worked out for the purposes of nonlocal field theories may be useful in this context. 

\acknowledgments

A large part of this work was done during my stay in Clausthal. I am indebted to the Alexander von Humboldt Foundation for funding and prof. H.-D. Doebner for hospitality and help.


\begin{references}
\bibitem{electron}Following the terminology of \cite{IZBB} we will
speak of negative electrons (negatons) and positive electrons
(positons).
\bibitem{tensor}We employ the representation where 
$$
A\otimes B=\left(
\begin{array}{ccc}
a_{11}B & a_{12}B &\dots\\
a_{21}B & a_{22}B &\dots\\
\vdots & \vdots & \ddots
\end{array}
\right).
$$
\bibitem{Noether}Construction of four-momentum and charge from the Noether theorem will be discussed elsewhere. In short, taking an appropriate Lagrangian one starts with field operators reducibly quantized in a finite volume and at the end performs the infinite volume limit. There is no need of normal ordering and no divergence occurs at any stage.
\end{references}
\end{document}